\documentclass{aa}
\usepackage{aabib99}
\usepackage{times}
\usepackage{graphicx}

\begin{document}

\thesaurus{03 
  (08.19.04; 
  08.19.5 \object{SN\,1991T}; 
  11.09.1 \object{NGC\,4527}; 
  12.04.3) 
}

\title{The brightness of SN\,1991T and the uniformity of decline-rate
  and colour corrected absolute magnitudes of supernovae Ia
  \thanks{Based on observations with the NASA/ESA Hubble Space
    Telescope, obtained at the Space Telescope Science Institute,
    which is operated by AURA, Inc., under NASA contract NAS 5-26555.}
  }

\titlerunning{The brightness of SN\,1991T}

\author{T.~Richtler\inst{1} \and J.B.~Jensen\inst{2}\thanks{Gemini
    Science Fellow} \and J.~Tonry\inst{3} \and B.~Barris\inst{3} \and
  G.~Drenkhahn\inst{4}}


\institute{Universidad de Concepci\'on, Departamento de F\'{\i}sica,
  Casilla 160-C, Concepci\'on, Chile, (tom@coma.cfm.udec.cl)\and
  Gemini Observatory, 670 N. A\`ohoku Place Hilo, HI 96720,
  (jjensen@gemini.edu)\and
  Institute for Astronomy, University of Hawaii 2680 Woodlawn Drive,
  Honolulu, HI 96822 \and
  Max-Planck-Institut f\"ur Astrophysik, Postfach 1317, 85741 Garching
  bei M\"unchen, Germany (georg@mpa-garching.mpg.de)}

\date{Received ... / Accepted ...}

\maketitle

\begin{abstract}
  We present a distance to \object{NGC\,4527}, the host galaxy of the
  type Ia \object{SN\,1991T}, measured by surface brightness
  fluctuations. This supernova has been labelled ``peculiar'' both on
  the grounds of its spectroscopic behaviour and its apparent
  overluminosity with respect to other supernovae. The distance
  modulus to \object{NGC\,4527} and thus to \object{SN\,1991T} is
  $30.26\pm0.09$. This relatively short distance largely removes the
  discrepancy with other Ia supernovae having similar light-curve
  characteristics and also removes the motivation for interpreting
  \object{SN\,1991T} as a super-Chandrasekhar explosion.  However, the
  reddening uncertainty results in significant uncertainty of the
  absolute magnitudes.
  
  We show further that, although the intrinsic absolute magnitudes
  among Ia supernovae are quite different, their brightness, corrected
  for decline-rate and colour, shows a high degree of uniformity,
  including underluminous events like \object{SN\,1991bg} and
  \object{SN\,1997cn}. In particular, the present data do not support
  a significant difference between corrected SNe Ia luminosities in late-type
 and early-type host galaxies. 

\end{abstract}
 
\section{Introduction}

A precondition for the successful use of Ia supernovae at high
redshifts for constraining cosmological models is the understanding of
low-redshift supernovae, in particular their luminosities and their
dependence on light-curve parameters such as the decline-rate
\cite{phillips:93}. Besides the decline rate, it is also the colour at
maximum, for which one has to correct \cite{tripp:98,drenkhahn:99}.
The Cal\'an-Tololo sample \cite{hamuy:96} and the Harvard sample
\cite{riess:99} demonstrated that most Ia supernovae, after
appropriate corrections, exhibit a surprisingly high degree of
uniformity in their maximum brightness.

A few supernovae, however, seem to deviate in the sense that they are
either too bright or too dim. Since at high redshifts, selection
effects towards the bright end of the supernovae luminosity
distribution are probably present, the apparently overluminous
supernovae are of particular interest.  Obviously, the distance to the
host galaxy is the key parameter.  For example, in the case of
\object{SN\,1994D} in \object{NGC\,4526}, it could be shown that this
apparently overluminous Ia event took place at a
shorter distance than initially assumed,
and thus its luminosity is consistent with other type-Ia SNe
\cite{drenkhahn:99}.

Another more extreme example of overluminosity is \object{SN\,1991T}
in \object{NGC\,4527}. Besides its allegedly high luminosity, it also
exhibited spectroscopic peculiarities.  Fisher et~al.
\cite*{fisher:99} discuss the present knowledge about
\object{SN\,1991T} and also offer an interpretation of these abnormal
spectral features.  Their analysis suggested that \object{SN\,1991T}
was a super-Chandrasekhar explosion following the merging of two white
dwarfs.

Although there is general agreement that \object{NGC\,4527} is located
on the near side of the Virgo cluster, there is no individual distance
determination besides a Tully-Fisher distance, where it is known that
large deviations can occur. Fisher et~al. \cite*{fisher:99} assumed
its distance to be equal to the distance of two neighboring spiral
galaxies, which also have been hosts for Ia SNe, \object{NGC\,4496}
(\object{SN\,1960F}), and \object{NGC\,4536} (\object{SN\,1981B}).

The aim of our paper is to present an individual distance to
\object{SN\,1991T}, based on surface brightness fluctuations in the
bulge of \object{NGC\,4527}, which is more accurate than what is
available in the literature.

Before we present our method of distance determination and its result,
we recall briefly what is known about the distance of this galaxy
group and the luminosities of the respective Ia SNe.

\subsection{What is known about the distances of \object{NGC\,4527} and 
  \object{NGC\,4536}?}

\object{NGC\,4536}, the host of \object{SN\,1981B}, has a more or less
well established distance, based on the PL-relation for Cepheids
observed by HST \cite{saha:96a,saha:96b}. However, the Cepheid data
are to a minor degree subject to interpretational manipulation,
depending on details of the data reduction and also of the selection
of Cepheids used for the distance determination.  Saha et~al.
\cite*{saha:96b}, for instance, quote a distance modulus to
\object{NGC\,4536} of $31.07$\,mag, while Gibson et~al.
\cite*{gibson:00} give $30.95\pm0.23$\,mag.
   
Tully-Fisher (TF) distances have been measured both for
\object{NGC\,4527} and for \object{NGC\,4536}. If we use the new TF
calibrations given by Sakai et al.  \cite*{sakai:00} and the TF data
given by Pierce \cite*{pierce:94}, we then get distance moduli
(averaged over $B$, $R$, $I$) of $30.78$ and $30.80$ for
\object{NGC\,4527} and \object{NGC\,4536}, respectively. Pierce
\cite*{pierce:94} himself quotes $30.6$ and $30.5$, respectively.  The
differences are consistent with the dispersion of the TF relation
which is about $0.3$--$0.4$\,mag.  In this sense, the TF data are in
agreement with the Cepheid distance of \object{NGC\,4536}. Moreover,
the TF distances cannot support any distance difference between
\object{NGC\,4527} and \object{NGC\,4536} and it appears reasonable to
assume a distance modulus of $30.95\pm0.23$\,mag for
\object{NGC\,4527} as well.


\section{Revised relations between SN Ia maximum brightness, 
  decline rate and colour}

It is now clear that the intrinsic maximum brightness of SNe Ia vary
by approximately 1\,mag. Using both the ``Cal\'an-Tololo'' sample
\cite{hamuy:96} and the Harvard sample \cite{riess:99}, we below
present handy relations between maximum brightness (extinction
corrected), decline rate and colour for the $B$, $V$, and $I$ bands.
We define the colour as the difference between the $B_\mathrm{max}$
and $V_\mathrm{max}$ maximum magnitudes, which is not necessarily the
$B-V$ colour at, say, the $B$-maximum. These relations are revised
versions of those given in Drenkhahn \& Richtler \cite*{drenkhahn:99},
which only included the Cal\'an-Tololo sample. A similar analysis,
based on the Cal\'an-Tololo sample alone, has been done by Tripp \&
Branch \cite*{tripp:99}.  A colour dependence is definitively present,
but it is difficult in the individual case to distinguish between an
intrinsic red colour and reddening. We also point out that only the
galactic foreground reddening can be accounted for, because internal
reddenings are not available. However, if reddening dominated the
colour dependence of SN luminosities, we would expect distinctly
larger colour coefficients in $B$ and $V$.
\begin{table*}[htbp]
  \caption{The table lists the coefficients according to our
    correction formula (\ref{eq:corr}) for the relation between
    brightness, decline rate, and colour.}
  \label{tab:coef}
  \begin{tabular}{cccccc}
    \hline
    Band & $Z$ & $b$ & $R$ & $\overline{\Delta m_{15}}$ & $\overline{B_\mathrm{max}-V_\mathrm{max}}$\\
    \hline
    $B$ & $-3.345\pm0.051$ & $-0.61\pm0.21$ & $-2.53\pm0.62$ & $1.237$ & $0.0$\\
    $V$ & $-3.356\pm0.041$ & $-0.66\pm0.18$ & $-1.62\pm0.53$ & $1.237$ & $0.0$\\
    $I$ & $-3.118\pm0.050$ & $-0.35\pm0.19$ & $-1.12\pm0.58$ & $1.230$ & $-0.017$\\
    \hline
  \end{tabular}
\end{table*}


The coefficients in Table~\ref{tab:coef} result from a
$\chi^2$-fit to the data of 41 (24 in the $I$ band) SN Ia from Hamuy
et~al. \cite*{hamuy:96} and Riess et~al. \cite*{riess:99}.
The listed values are the best fits to the relation
\begin{equation}
  \label{eq:corr}
  \begin{array}{l}
  m_\mathrm{max}
  +b\cdot(\Delta m_{15}-\overline{\Delta m_{15}})\\\quad
  +R\cdot(B_\mathrm{max}-V_\mathrm{max}-(\overline{B_\mathrm{max}-V_\mathrm{max}}))\\\quad
  = 5\cdot\log cz + Z
  \end{array}
\end{equation}
with $m_\mathrm{max}$ being the foreground extinction corrected
magnitudes in $B$, $V$ and $I$, $b$ the decline-rate coefficient, $R$
the colour coefficient, and $Z$ the zero-point of the Hubble diagram
defined by the merging of both SNe samples.  For the determination of
the foreground extinction we used the dust maps of Schlegel et~al.
\cite*{schlegel:98}.

Table~\ref{mags} lists the photometric parameters for our SNe taken
from Hamuy et~al. \cite*{hamuy:96}. The reddening values and their
uncertainties are those quoted by Fisher et~al. \cite*{fisher:99}.
 
Using the correction terms for colour and decline-rate, and the
Cepheid distance modulus for \object{NGC\,4536}, one arrives at the
corrected absolute maximum magnitudes given in Table~\ref{absmag}.

The Cal\'an-Tololo and the Harvard sample demonstrated that there is
no observable offset between the corrected absolute magnitudes of SNe
Ia in early-type and late-type galaxies. Within the uncertainty
limits, \object{SN\,1981B} matches the SNe Ia in the Fornax cluster
and also to some of the Ia's in other spiral galaxies, although
uncertain photometry and uncertain extinction make the comparison more
difficult \cite{drenkhahn:99}. However, the brightness of
\object{SN\,1991T} is inconsistent with that of \object{SN\,1981B} or
other supernovae, if the Cepheid distance of \object{NGC\,4536} is
also adopted for \object{NGC\,4527}.  Table~\ref{absmag} demonstrates
that this discrepancy is not very dependent on the adopted reddening,
although the error bars overlap, almost exclusively as a result of the
reddening uncertainty (see Fisher et~al. \cite*{fisher:99} for a
compilation of the relevant literature).

\begin{table*}
  \caption{This table lists the photometric parameters for 
    \object{SN\,1991T} and \object{SN\,1981B} SNe according to Hamuy
    et~al. \protect\cite*{hamuy:96}. The $B$, $V$, $I$ magnitudes are
    not extinction corrected.  The reddening values and their
    uncertainties are those quoted by Fisher
    et~al. \protect\cite*{fisher:99} and the uncertainty of the
    reddening of \object{SN\,1991T} reflects the variety of literature values.}
  \label{mags}
  \begin{tabular}{lllllll}
    \hline
    Host Galaxy & SN    & $B_\mathrm{max}$ & $V_\mathrm{max}$ & $I_\mathrm{max}$ & $E(B-V)$      & $\Delta m_{15}$\\
    \hline
    NGC\,4527   & 1991T & $11.69\pm0.03$   & $11.51\pm0.03$   & $11.62\pm0.03$   & $0.20\pm0.10$ & $0.94\pm0.05$\\ 
    NGC\,4536   & 1981B & $12.03\pm0.03$   & $11.93\pm0.03$   & ---              & $0.10\pm0.05$ & $1.10\pm0.05$\\
    \hline
  \end{tabular}
\end{table*}
 
\begin{table*}
  \caption{
    This table lists the corrected absolute maximum magnitudes of the
    SNe according to (\ref{eq:corr}) and a distance modulus of $30.95$,
    corresponding to the
     Cepheid distance of NGC\,4536.  The photometric parameters are
    taken from Table~\ref{mags} and are extinction corrected with
    $A_B/E(B-V)=4.1$, $A_V/E(B-V)=3.1$ and $A_I/E(B-V)=1.5$
    \protect\cite{rieke:85}.  To illustrate the effect of a different
    reddening, we also give for SN\,1991T the values for a reddening
    of $E(B-V)=0.13$ which has been advocated by Phillips
    et~al. \protect\cite*{phillips:92}.}
  \label{absmag}
  \begin{tabular}{llcccc}
    \hline
    Host Galaxy & SN                    & $\mu$          & $M_{B,\mathrm{max}}$ & $M_{V,\mathrm{max}}$ & $M_{I,\mathrm{max}}$\\
    \hline
    NGC\,4527   & 1991T ($E(B-V)=0.20$) & $30.95\pm0.09$ & $-19.87\pm0.51$ & $-19.85\pm0.38$ & $-19.64\pm0.17$\\  
    NGC\,4527   & 1991T ($E(B-V)=0.13$) & $30.95\pm0.09$ & $-19.75\pm0.28$ & $-19.74\pm0.21$ & $-19.53\pm0.11$\\      
    NGC\,4536   & 1981B                 & $30.95\pm0.23$ & $-19.26\pm0.26$ & $-19.25\pm0.20$ & --- \\ 
    \hline
  \end{tabular}
\end{table*}


\newcommand{\Mbar}{\bar{M}}
\newcommand{\mbar}{\bar{m}}

\section{The distance to NGC\,4527 from surface brightness fluctuations}

We measured a new and accurate distance to NGC\,4527 using the near-IR
surface brightness fluctuation technique.  Two 128-second images of
NGC\,4527 taken with NICMOS on the Hubble Space Telescope were
recovered from the archive.  These images were taken using the NIC2
camera using the F160W filter, and have a spatial resolution of
$0\farcs13$ FWHM.  The excellent resolution and very low background
(approximately 18.9\,mag/$\sq\arcsec$ 
at F160W) make detection of surface brightness fluctuations
straight-forward in the bulge of NGC\,4527.

The calibration and techniques for determining IR SBF distances using
NIC2 in the F160W bandpass were developed by Jensen et~al.
\cite*{jensen:00}.  This calibration relies on direct measurements of
SBFs, both in the I and F160W bands, to nearby spirals with
well-determined Cepheid distances \cite{tonry:00,jensen:00}.  The
absolute fluctuation magnitude calibration $\Mbar=-4.86\pm0.05$
(statistical uncertainty only) presented by Jensen et~al.
\cite*{jensen:00} is essentially constant with galaxy colour redward
of $(V{-}I)=1.16$.  NGC\,4527 has a colour of $(V{-}I)=1.23\pm0.03$
(J.~Tonry and the optical SBF team, private communication) in the
region of the galaxy where the SBF analysis was performed.  We adopted
the Galactic extinction for NGC\,4527 from Schlegel, Finkbeiner, \&
Davis \cite*{schlegel:98} $A_B=0.095$ mag, implying a correction of
only 0.01\,mag at F160W.  Jensen et~al.'s calibration also showed
excellent agreement between giant ellipticals and the bulges of spiral
galaxies, thereby confirming that the calibration applies equally well
to the old, smoothly-distributed stellar populations in the centers of
spirals as to the giant elliptical galaxies traditionally targeted for
SBF studies.

The methods for measuring SBFs are described in several papers (see
Blakeslee et~al. \cite*{blake:99} for a review); the infrared
techniques used here are are described by Jensen, Tonry, \& Luppino
\cite*{jensen:98} and Jensen et al. \cite*{jensen:00}.  Basically, the
discrete nature of stars in galaxies leads to variations in the number
of stars imaged by each pixel in the detector because of Poisson
fluctuations.  The fluctuations in surface brightness are
distance-dependent: distant galaxies look smooth compared to nearby
ones because the number of luminous stars imaged by each pixel is
larger.  The variations are convolved with the point-spread function
(PSF) of the telescope, in this case the diffraction pattern of HST as
imaged by the NIC2 camera.  In practice we observe the variations in
luminosity from pixel to pixel rather than variations in the number of
stars, so surface brightness fluctuations are dominated by the
brightest stars in the galaxy.  Because the most luminous stars are
red giants in old populations, SBFs are approximately 30 times
brighter at F160W than in the optical $I$ band.

\begin{figure}
  \includegraphics[bb=36 266 573 526,angle=00,width=1.0\hsize,clip]{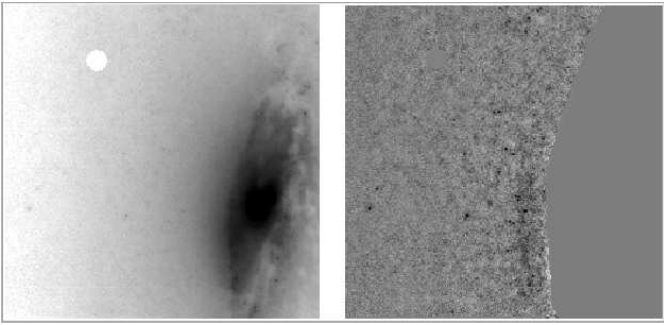}
  \caption{The NIC2 F160W image of NGC\,4527 is shown using a square-root
    grayscale on the left.  The same image with the smooth galaxy
    profile subtracted is shown on the right, using a linear stretch.
    Point sources apparent in the galaxy-subtracted image were masked
    prior to measuring the spatial power spectrum.  The field of view
    is $19\farcs2$ across.}
  \label{Figure1}
\end{figure}

The initial image processing proceeded with the galaxy images being
flattened and combined after a small bias correction was applied to
remove mismatches between detector quadrants.  During each of the two
exposures, the NIC2 array was read multiple times to allow for the
temporal rejection of cosmic rays.  Additional cosmic rays were
corrected when combining the two images.  The results are shown in the
left panel of Fig.~\ref{Figure1}.  The inner few arcsec of NGC\,4527
are choked with dust lanes in the disk, and these were masked prior to
proceeding with the SBF analysis.  Point sources were also identified
and masked.  A smooth model of the galaxy luminosity profile was then
constructed and subtracted.  An image of NGC\,4527 with the galaxy
subtracted, showing the stellar SBFs, is shown in the right panel of
Fig.~\ref{Figure1}.

\begin{figure}
  \includegraphics[bb=0 0 500 401,angle=00,width=1.0\hsize,clip]{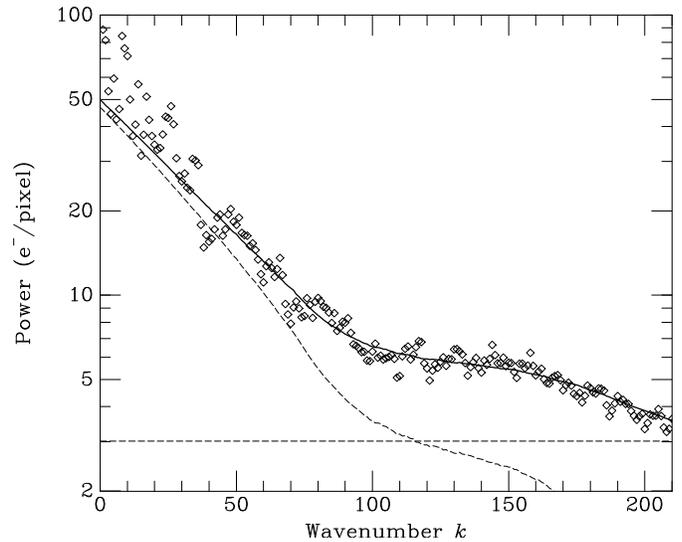}
  \caption{The azimuthally-averaged spatial power spectrum is shown
    for the galaxy-subtracted image with point-sources removed.  The
    solid line is the sum of a white-noise (constant) component and
    the scaled PSF power spectrum (both shown with dashed lines).  The
    scaling of the normalized PSF is the power at $k=0$, and is the
    surface brightness fluctuations flux from which the fluctuation
    magnitude is computed.}
  \label{Figure2}
\end{figure}

The spatial power spectrum was constructed and fitted to a linear
combination of the scaled PSF power spectrum and a constant,
white-noise background.  The two components are plotted using dashed
lines in Fig.~\ref{Figure2}, and the solid line is their sum.  The
``fluctuation power'' (in electrons/pixel) is the scale factor by
which the power spectrum of the normalized PSF must be multiplied to
match the data over the range in wavenumbers where the PSF is
dominant.  The fit shown in Fig.~\ref{Figure2} was constructed for
wavenumbers between $k\,{=}\,40$ and 100, corresponding to spatial
scales between $0\farcs32$ and $0\farcs96$, where the PSF component
dominates.  Lower wavenumbers are not used because the power spectrum
is contaminated by extra power from large-scale residuals of the
galaxy subtraction; at wavenumbers higher than 100 the white-noise
component dominates.  The fit between $k\,{=}\,40$ and 100 is
excellent across the entire power spectrum, and yields a fluctuation
power of $46.8\pm2.6$ electrons/pixel in the 256\,s exposure.  The
fluctuation power is 15.6 times larger than the white-noise component
(effectively the S/N ratio of the observation).

Once the fluctuation power was determined, the apparent fluctuation
magnitude was easily computed and the distance modulus determined
using the Jensen et~al. \cite*{jensen:00} calibration.  The apparent
fluctuation magnitude of NGC\,4527 is $\mbar=25.40\pm0.07$, where the
uncertainty is the statistical uncertainty arising from the power
spectrum fit (0.054\,mag), the uncertainty in the PSF normalization
(0.061\,mag), and the uncertainty in the sky background subtraction
(0.028\,mag), all added in quadrature.  As the PSF and sky were not
measured explicitly for the observations of NGC\,4527, we used the
best values and uncertainties from the library of values collected by
Jensen et~al. \cite*{jensen:00}.  Subtracting the absolute fluctuation
magnitude $\Mbar$ yields a distance modulus of $(m{-}M)=30.26\pm0.09$,
corresponding to a distance (good to 5\%) of $11.3\pm0.50$\,Mpc.
NGC\,4527 is significantly closer than the bulk of the Virgo cluster.

The IR SBF calibration presented by Jensen et~al. \cite*{jensen:00} is
derived from Cepheid distances to a handful of nearby spirals
\cite{ferrarese:00}, and is hence subject to any systematic
uncertainties that affect the Cepheid distance scale generally (such
as the distance to the LMC or any period-luminosity dependence on
metallicity, for example).  The distance modulus uncertainty for
NGC\,4527 quoted here does not include the additional systematic
uncertainty from the Cepheid distance (estimated to be $0.16$\,mag)
scale beyond that incurred by linking IR SBFs to the Cepheid
distances.



\section{The comparison with other SNe Ia}

It is now interesting to compare SN\,1991T with other Ia SNe in spiral
and early-type galaxies, for which more or less accurate distances are
known from either Cepheids, from globular cluster luminosity
functions, or from surface brightness fluctuations.  If we keep the
coefficients for decline-rate and colour correction fixed and
distinguish between late-type and early-type host galaxies, the
resulting zero-points are for $B$, $V$, $I$, respectively (the number
of SNe involved are in parentheses): $-3.34$ (14), $-3.32$ (14),
$-3.04$ (9) for early-type host galaxies, and $-3.29$ (29), $-3.33$
(29), $-3.11$ (16) for late-type host galaxies.  It is apparent that
there is no significant difference in the corrected absolute
magnitudes between SNe Ia in early-type and late-type hosts. We
therefore expect the same for SNe in galaxies whose distances are
known.

\subsection{SNe in late-type galaxies}

Gibson et al. \cite*{gibson:00} give a compilation of parameters for
the 8 SNe Ia in late-type galaxies, which have measured Cepheid
distances.  We adopt them here since they are also the basis for the
SFB calibration using Cepheids. To this list, we add SN\,1991T. The
reddenings used for the extinction calculation are adopted from
Phillips et al. \cite*{phillips:99}.  With these data and the
correction terms for decline rate and colour presented above, we
obtain the corrected absolute magnitudes compiled in
Table~\ref{complate}. The uncertainties in the table (in mag) result
from a formal error propagation of the different parameters involved.
The formally resulting Hubble constant values are also given as the
mean of the individual Hubble constant values in the different
filters.

Uncertainties in the distances do not necessarily dominate the total
uncertainties.  Uncertain reddening values and hence uncertain maximum
colours make a significant contribution in the case of reddened SNe.
This is illustrated in the case of SN\,1991T, where the two different
reddening values from the literature are adopted (see
Table~\ref{absmag}). Moreover, in some cases the adopted light curve
parameters are more a matter of historical reconstruction than of
exact measurements. This is the case for \object{SN\,1937C},
\object{SN\,1960F}, and \object{SN\,1974G} (see van den Bergh
\cite*{vandenbergh:96} for the relevant literature).

If we skip \object{SN\,1991T} because of the large reddening
uncertainty, and \object{SN\,1937C}, \object{SN\,1960F}, and
\object{SN\,1974G} from the SNe with Cepheid distances, we get
$-19.15\pm0.15$, $-19.15\pm0.13$, $-18.93\pm0.17$\,mag as the weighted
means (weight factors are the inverse photometric uncertainties) of
the absolute magnitudes for $B$, $V$, and $I$, respectively. The
corresponding Hubble constant values are $68.9\pm4.7$, $69.3\pm4.1$,
$68.7\pm4.1$\,km/s/Mpc.

Including all SNe in Table~\ref{complate}, and adopting for
\object{SN\,1991T} the higher reddening value of $E(B-V)=0.2$\,mag, we
get $-19.25\pm0.13$, $-19.25\pm0.11$, $-18.93\pm0.13$. The
corresponding Hubble constant values are $65.8\pm3.8$, $65.9 \pm3.3$,
$68.7\pm4.1$\,km/s/Mpc.

We thus face the unsatisfactory situation that in spite of the huge
effort which has gone into the determination of distances, the
resulting Hubble constant is still influenced by the ``freedom'' of
selecting the ``right'' SNe.

\begin{table*}
  \caption{
    This table lists the corrected absolute maximum magnitudes of 8
    SNe Ia, whose distances have been measured by Cepheids, together
    with \object{SN\,1991T}, for which now the new distance modulus of 30.26
    has been used. The distances and light curve parameters are
    adopted from Gibson et al. \protect\cite*{gibson:00}, the
    reddening values, except those for \object{SN\,1991T}, are from Suntzeff et
    al. \protect\cite*{suntzeff:99}. The uncertainties follow from an
    error propagation of the various parameters involved. Systematic
    distance scale uncertainties are not accounted for. The
    corresponding Hubble constants are the average values (not
    weighted) of the respective band values.
    }
  \label{complate} 
  \begin{tabular}{lllllll}
    \hline
    Host Galaxy (type) &  SN & $M_{B,\mathrm{max}}$ & $M_{V,\mathrm{max}}$ & $M_{I,\mathrm{max}}$ & $H_0$\\
    \hline
    NGC\,4527 (Sab)     & 1991T ($E(B-V)=0.20$) & $-19.18\pm0.52$ & $-19.16\pm0.39$ & $-18.95\pm0.18$ & $68.6\pm11.6$\\
    NGC\,4527 (Sab)     & 1991T ($E(B-V)=0.13$) & $-19.06\pm0.29$ & $-19.05\pm0.22$ & $-18.84\pm0.18$ & $72.2\pm7.3$\\
    NGC\,4639 (Sab)     & 1990N                 & $-19.09\pm0.28$ & $-19.09\pm0.27$ & $-18.86\pm0.25$ & $71.3\pm8.9$\\
    NGC\,4536 (Sab)     & 1981B                 & $-19.26\pm0.27$ & $-19.25\pm0.21$ & ---             & $66.1\pm7.4$\\
    NGC\,3627 (Sb)      & 1989B                 & $-19.19\pm0.41$ & $-19.19\pm0.37$ & ---             & $68.0\pm12.4$\\
    NGC\,3368 (Sab)     & 1998bu                & $-19.28\pm0.41$ & $-19.27\pm0.35$ & $-19.09\pm0.28$ & $65.0\pm10.5$\\
    NGC\,5253 (Im pec)  & 1972E                 & $-18.90\pm0.40$ & $-18.88\pm0.37$ & $-18.81\pm0.34$ & $76.4\pm13.1$\\
    IC\,4182 (SAm)      & 1937C                 & $-19.29\pm0.37$ & $-19.27\pm0.33$ & ---             & $65.5\pm10.6$\\
    NGC\,4496A (SB)     & 1960F                 & $-19.55\pm0.40$ & $-19.65\pm0.37$ & ---             & $56.4\pm10.2$\\
    NGC\,4414 (SAc)     & 1974G                 & $-19.58\pm0.47$ & $-19.57\pm0.39$ & ---             & $57.1\pm11.3$\\
    \hline
  \end{tabular}
\end{table*}

\subsection{SNe Ia in early-type galaxies}

SNe Ia in S0 or elliptical galaxies are less affected by extinction.
There exist some SNe with good photometry in nearby early-type
galaxies (Tab.~\ref{compearly}).  The Fornax cluster plays a particularly important role by hosting
 three SNe Ia: \object{SN\,1981D} and
\object{SN\,1980N} in \object{NGC\,1316}, and \object{SN\,1992A} in
\object{NGC\,1380}. Of these, the light curve of \object{SN\,1981D}
has a bad photometry and we skip it here. Then there is
\object{SN\,1994D} in \object{NGC\,4526} (Drenkhahn \& Richtler
\cite*{drenkhahn:99}) and \object{SN\,1991bg} in \object{NGC\,4374}.
The latter has a reputation as a strongly underluminous event.

We adopt the light-curve parameters from the following sources:
\object{SN\,1980N}, \object{SN\,1991bg}, and \object{SN\,1992A} from
Hamuy et al. \cite*{hamuy:96}, \object{SN\,1994D} from Drenkhahn \&
Richtler \cite*{drenkhahn:99}.  The distance of \object{SN\,1994D} has
been determined by the method of globular cluster luminosity functions
and we adopt it also from Drenkhahn \& Richtler \cite*{drenkhahn:99}
($30.4\pm0.3$).  Della Valle et al. \cite*{dellavalle:98} applied this
method to \object{SN\,1992A} and got $31.3\pm0.16$, which is in very
good agreement with the distance of Fornax galaxies determined by
surface brightness fluctuations (Jensen et al. \cite*{jensen:98}) and
with two of the three late-type galaxies, for which Cepheid distances
are available (Ferrarese et al. \cite*{ferrarese:00}).  There exists
no published individual distance yet for \object{NGC\,1316}, but
preliminary results from the analysis of its globular cluster system
(G\'omez et al., in preparation) indicates very good agreement with SN
1992A and we adopt $30.35\pm0.25$. For \object{NGC\,4374}, we take the
distance modulus quoted by Ferrarese et al. \cite*{ferrarese:00}
($31.2\pm0.1$), which is in very good agreement with the recent SFB modulus
of ($31.17\pm0.1$) by Neilsen \& Tsvetanov \cite*{neilsen:00}.

\begin{table*}
  \caption{
    This table lists the corrected absolute maximum magnitudes of SNe
    Ia in early-type galaxies. Refer the text for data sources.}
  \label{compearly} 
  \begin{tabular}{lllllll}
    \hline
    Host Galaxy (type) & SN & $M_{B,\mathrm{max}}$ & $M_{V,\mathrm{max}}$ & $M_{I,\mathrm{max}}$ & $H_0$\\
    \hline
    NGC\,1380 (S0)      & 1992A  & $-19.01\pm0.29$ & $-19.02\pm0.27$ & $-18.58\pm0.25$ & $76.2\pm9.7$\\
    NGC\,1316 (S0pec)   & 1980N  & $-19.05\pm0.30$ & $-19.05\pm0.27$ & $-18.68\pm0.25$ & $74.2\pm9.5$\\
    NGC\,4526 (S0)      & 1994D  & $-18.76\pm0.35$ & $-18.72\pm0.34$ & $-18.46\pm0.31$ & $81.5\pm12.2$\\
    NGC\,4374 (E)       & 1991bg & $-19.03\pm0.61$ & $-19.10\pm0.49$ & $-18.75\pm0.12$ & $73.1\pm13.7$\\
    \hline
  \end{tabular}
\end{table*}

The weighted average values (weight factors are the inverse
uncertainties) of the absolute, corrected magnitudes are
$-18.96\pm0.18$, $-18.97\pm0.16$, $-18.66\pm0.10$ for $B$, $V$, $I$,
respectively.  The corresponding Hubble constant values are
$75.3\pm6.2$, $75.4\pm5.4$, $78.1\pm3.8$\,km/s/Mpc.

Besides \object{SN\,1991bg}, another strongly underluminous event was
\object{SN\,1997cn} in the elliptical galaxy \object{NGC\,5490}
\cite{turatto:98}. No individual distance determination is known, but
the galaxy should be well in the Hubble-flow (heliocentric recession
velocity 5008\,km/s).  Therefore, a consistency check is possible by
assuming a Hubble constant of 70\,km/s/Mpc and correspondingly a
distance modulus of 34.2\,mag. Given the light-curve data by Turatto
et~al.  \cite*{turatto:98} and assuming zero extinction, we get for
the corrected absolute magnitudes in $B$, $V$, $I$ $-19.02\pm0.57$,
$-19.11\pm0.45$, $-18.85\pm0.11$.  Thus even such a red SN like
\object{SN\,1997cn} with $B_\mathrm{max}-V_\mathrm{max}=0.65$ is
consistent with the other SNe. It is striking that in spite of the
large errors, especially in B, caused by the large and uncertain
colour corrections, the absolute magnitudes of \object{SN\,1997cn} and
\object{SN\,1991bg} agree so well with those of the other SNe. This is
perhaps an indication that due to the small number of red SNe, the
error of the colour correction is overestimated.

\section{Discussion}

The formal brightness differences between the SNe in
Tabs.~\ref{complate} and \ref{compearly} are $-0.30\pm0.25$,
$-0.29\pm0.25$, $-0.31\pm0.17$ for $B$, $V$, $I$, respectively, if all
SNe are taken ($E(B-V)=0.2$ in case of \object{SN\,1991T}). If
\object{SN\,1991T}, \object{SN\,1960F}, and \object{SN\,1974G} are
skipped, the differences reduce to $-0.21\pm0.2$, $-0.19\pm0.22$, and
$-0.3\pm0.2$. Most SNe Ia in late-type galaxies probably have
different progenitors from SNe Ia in early-type galaxies, as the
distribution of decline-rates suggests: we find only fast decliners in
early-type galaxies, but the entire range of decline-rates in
late-type galaxies (e.g. Richtler et al. \cite*{richtler:00b}), which,
however, are dominated by slow-decliners. 

Also Tripp \& Branch \cite*{tripp:99} got a marginally smaller Hubble 
constant if they select only spirals from the Cal\'an-Tololo sample in
combination with Cepheid calibrated host galaxies.  
However, the corresponding brightness difference is within the statistical
uncertainty. Since the combined 
Cal\'an-Tololo and Harvard samples show that the decline-rate and
colour corrections completely account for the intrinsic brightness
difference, we are apparently forced to draw the conclusion that any
remaining difference after correction is due to a systematic
difference in the distances to early- and late-type host galaxies.
But even that is not entirely clear. The Fornax cluster distance is dominant
for the average Ia brightness in early-type host galaxies. A distance
modulus of the Fornax cluster of $31.60$\,mag, as now advocated by
Ferrarese et al. \cite*{ferrarese:00}, would brighten the two Fornax
Ia's in Table \ref{compearly} by about $0.2$\,mag, which would cause
the brightness difference almost to vanish.

On the other hand, the case for $31.35$\,mag as the Fornax distance
modulus is strong (Jensen et al. \cite*{jensen:98}, Richtler et al.
\cite*{richtler:00a}). Moreover, two of the three Fornax spirals with
Cepheid distances from Ferrarese et al.  are quoted with distance
moduli $31.43\pm0.07$ and $31.39\pm0.10$ and thus agree very well with
the distance based on surface brightness fluctuations and globular
cluster luminosity functions. The Hubble constant values from Tab.\ref{compearly} also fit well to the value of 77 km/s/Mpc, emerging from the SFB survey
by Tonry et al. \cite*{tonry:00}. But since some of the supernovae with Cepheid
distances still agree  within the uncertainties, it would be premature to
claim a difference between SFB distances and Cepheid distances.   
 
Based on the sample of spiral host galaxies, Tripp \& Branch \cite*{tripp:99}
derive $M_B = -19.46$ as the colour and decline-rate corrected blue
luminosity for SNe Ia. One notes, however, that the mean difference
between their adopted distance moduli plus apparent brightness and our
values is already $0.3$\,mag. This demonstrates how uncertain the SN
Ia brightness in spiral galaxies still is. A deeper discussion, particularly
on the Hubble constant and the zero-point uncertainties of the distance
scale, is beyond
the scope of our paper and we refer the reader to Branch \cite*{branch:98},
Gibson et al. \cite*{gibson:00}, Tonry et al. \cite*{tonry:00} and references therein. 

Given the above arguments, the present data on supernovae absolute magnitudes does at
best marginally support a small difference in the absolute corrected
magnitudes of SNe in early-type and late-type galaxies, which,
however, is of the order of present uncertainties in the respective absolute
distance scales.

We further recall that systematic uncertainties in the distance scale
of Cepheids and globular cluster luminosity functions have not yet
been included.  For example, all distance moduli based on Cepheids use
$18.50$\,mag as the LMC distance modulus, whereas values between
$18.2$ and $18.7$\,mag appear in the literature, e.g. Walker
\cite*{walker:99}.

One can only conclude that progress regarding a further decrease of
the uncertainty of the Hubble constant must come from both sources: we
need more \emph{well} observed SNe Ia in nearby spirals and early-type
galaxies \emph{and} a more accurate zero-point determination of the
distance scale.

\section{Summary and Conclusions}

By applying the method of surface brightness fluctuations to the bulge
of \object{NGC\,4527}, we measured an accurate distance to this
galaxy, which hosted \object{SN\,1991T}. Our result for the distance
modulus is $30.26\pm0.09$.  With this distance, the absolute maximum
magnitudes of \object{SN\,1991T}, corrected for decline rate and
colour, in $B$, $V$, $I$ are $-19.18\pm0.52$, $-19.16\pm0.39$, and
$-18.95\pm0.18$, respectively (adopting a reddening of
$E(B-V)=0.2\pm0.1$.  Although the uncertainty is large, mainly caused
by the uncertain reddening, there is no reason to conclude that
\object{SN\,1991T} has a peculiarly high luminosity, as the comparison
with other SNe shows. Thus the model of a super-Chandrasekhar
explosion advanced by Fisher et~al.  \cite*{fisher:99} can not be
motivated by the new distance.
 
Since also underluminous events like \object{SN\,1991bg} and
\object{SN\,1997cn} are consistent with the corrected absolute
magnitudes of the other SNe, there is now no single SN known, which
deviates strikingly, in accordance with the homogeneous samples of the
Cal\'an-Tololo and the Harvard sample. Since there is evidence that
SNe Ia in late-type and early-type galaxies stem from different parent
populations, (see Leibundgut \cite*{leibundgut:00} for a general
review on SNe Ia) this may indicate that there is no evolutionary
effect as far as the corrected absolute magnitudes are concerned. This
is encouraging for the use of SNe Ia in cosmological contexts.

\acknowledgements{We greatly appreciate the help of the NICMOS team,
  which provided advice, software and the photometric zero point used
  in the SBF analysis.  We also acknowledge the helpful contribution
  of the optical colour data by the Optical SBF team (J. Tonry, J.
   Blakeslee, E. Ajhar, and A. Dressler).
The NICMOS GTO team was supported by NASA grant NAG 5-3042.  
J. Jensen acknowledges the support of the Gemini Observatory,
which is operated by the Association for Research in Astronomy, Inc.,
under a cooperative agreement with the National Science Foundation on
behalf of the Gemini partnership: the National Science Foundation
(United States), the Particle Physics and Astronomy Research Council
(United Kingdom), the National Research Council (Canada),  CONICYT
(Chile), the Australian Research Council (Australia), CNPq (Brazil)
and CONICET (Argentina).  J. Tonry and B. Barris were supported in part by 
NASA grant GO-07453.0196A. }

\bibliography{aamnem99,ms10183}
\bibliographystyle{aabib99}

\end{document}